\documentclass[12pt]{article}
\usepackage{amsmath}
\usepackage{eucal}
\usepackage{amsfonts}
\usepackage{amssymb}
\usepackage{amsthm}
\usepackage{amscd}

\newcommand{\hepth}[1]{{\tt hep-th/#1}}
\newcommand{\plb}[3]{Phys.Lett. {\bf B#1} (#2) #3}

\title{\bf Integrable models with boundaries and defects}
\author{E. Corrigan\footnote{
email: ec9@york.ac.uk}\\ \\Department of Mathematics,\\University of York, UK\\
}
\date{}

\begin{document}

\maketitle

\begin{quote}Two lectures given at
the UK-Japan Winter School on Geometry and Analysis Towards Quantum Theory,
Durham, January 2004.\end{quote}

\section{Introduction}

Over the past forty years there has been much effort devoted to
understanding various aspects of one-dimensional (that is, one
space - one time, generally denoted by $x$ and $t$, respectively)
field theories - both from the classical and
quantum points of view - and their applications in a variety of
contexts. For the most part, attention has been paid to models in
the bulk (that is, defined without restriction over the whole space
$-\infty <x < \infty$), or models defined on a circle (that is,
periodic in $x$). In these two lectures it is intended to describe
in a quite straightforward manner some of the ideas and complications
when boundaries of various kinds are introduced.

For simplicity, two particular cases will be considered. Firstly,
a free massive field will be used to illustrate the basic ideas;
secondly, the affine Toda series of models will be used to illustrate
the additional features coming into play when boundaries and their
associated boundary conditions are required to preserve integrability.
It will appear that a number of traditional ideas will need to be
adjusted and extended to accommodate this new situation.

Since the affine Toda field theories as bulk models have some
significance in a geometrical context (for example, see \cite{rs}),
it may be fruitful to wonder
about the role their integrable boundary conditions might play.

\section{Bulk theories}

The simplest of all bulk field theories is a single, free, massive
scalar field described classically by the Klein-Gordon equation
\begin{equation}\label{KG}
    ({\partial }^2+m^2)\phi =0,\qquad \partial^2 \equiv \partial_t^2-\partial_x^2.
\end{equation}
If the mass parameter $m$ is zero the field theory is conformal. From a quantum
perspective this field describes noninteracting particles of mass $m\hbar$ (henceforth
$\hbar$ will be set to unity), whereas from a classical point of view \eqref{KG} is
completely solvable using a Fourier transform technique.

Considerably more interesting is the sine-Gordon model whose field equation is
\begin{equation}\label{sG}
 {\partial }^2\phi =-\frac{m^2}{\beta}\sin\beta \phi,
\end{equation}
where $m$ provides the mass scale, as before, and $\beta$ is the bulk coupling.
Actually, the latter is only relevant in the quantum field theory since it can be removed from
the classical field equation by rescaling the field. If the field $\phi$ is regarded as `small'
then the linearised version of \eqref{sG} is \eqref{KG}. However, the nonlinearity
plays a fundamentally crucial role since it permits the existence of solitons, antisolitons
and breathers (see \cite{scm,l} for  reviews on solitons in general, and \cite{ft}
for a treatise on integrability), which means the quantum field theory is much more
interesting. The spectrum of particles contains the soliton, the antisoliton, and
a collection of breathers whose precise number is coupling dependent. These are not
free particles and their mutual scattering properties are fascinating (see, for
example, the classic
paper \cite{zz}).
If the sine is replaced by a hyperbolic sine on the right hand side of \eqref{sG}
the spectrum of states simplifies dramatically (essentially because the $2\pi$-periodicity
in the field is lost) and there is a single scalar particle, though it is not free
but scatters with other similar particles.

The sine(sinh)-Gordon model is the first of a series of models, the affine Toda models,
each of which is defined in terms of an extended root system. More details on this will
be given later but for now consideration of boundaries of various kinds will be restricted
to fields of type \eqref{KG} or \eqref{sG}.  For a fairly recent review, see \cite{c}.

\section{One boundary}
Consider a free massive scalar field restricted to the left half-line $(x < 0)$ by a linear
boundary condition at $x=0$. Its equation of motion and boundary condition are as follows:
 \begin{equation}\label{KGhalfline}
  \begin{cases}
   ({\partial }^2+m^2)\phi =0 , & \text{ $(x < 0)$ }\\
    \partial _x\phi =-\lambda \phi , & \text{ $(x = 0),$ }
  \end{cases}
 \end{equation}
which follow formally using the Euler-Lagrange equations applied to the Lagrangian
density
\begin{equation}\label{KGhalflinelagrangian}
{\cal L}= \theta(-x)\frac{1}{2}\left( (\partial\phi)^2 - m^2\phi^2\right)-\delta(x)\frac{1}{2}
\lambda \phi^2,
\end{equation}
where the last term plays the role of a boundary potential. Note, the boundary condition is
assumed to be homogeneous with respect to the field $\phi$; a slightly more general possibility
would add a constant to the right hand side of the expression for $\partial_x\phi$ and the
effect of that would be to force the lowest energy solution (with $\lambda >0$) to be $x$-dependent
rather than the constant $\phi=0$.

The pair of equations \eqref{KGhalfline} is solvable and a wave travelling towards the
boundary at $x=0$ will be reflected. Thus
\begin{equation}\label{KGreflection}
\phi(x,t) =e^{-i\omega t}\left(e^{ikt}+R(k)e^{-ikx}\right) + cc, \quad \omega ^2=k^2+m^2,
\end{equation}
where $R(k)$ is a `reflection' factor and `cc' denotes complex conjugate  (necessary to
keep $\phi$ real).
Clearly, $R$ can be calculated using the boundary condition to find
\begin{equation}\label{KGR}
R(k)=\displaystyle \frac {ik+\lambda }{ik-\lambda }.
\end{equation}

On the other hand, if the field $\phi$ is not free, but rather satisfies an
equation such as \eqref{sG}, then it is  natural to ask what
boundary conditions (if any) are compatible with integrability. In those circumstances,
the generalisation of \eqref{KGhalfline} is
\begin{eqnarray}\label{shGhalfline}
\nonumber{\partial }^2\phi &=-\frac{m ^2}{\beta}\sinh\beta \phi ,  \quad (x < 0) \\
       \partial _x\phi &=-\frac{\partial B}{\partial \phi } , \ \quad\quad\quad\quad (x = 0),
 \end{eqnarray}
following in the usual way from the Lagrangian
\begin{equation}\label{shGhalflinelagrangian}
\mathcal{L}=\theta (-x)\left(\frac{1}{2} {(\partial \phi )}^2 -{\cal V}(\phi )\right) -
\delta (x){\cal B}(\phi ),
\end{equation}
where ${\cal B}(\phi)$ is the boundary potential. For reasons which will be described
in more detail later, integrability requires
\begin{align}\label{sinhGpotentials}
   {\cal B} (\phi ) &= \frac{2}{\beta^2}\left(b_1e^{\beta\phi/2} + b_0e^{-\beta\phi /2}\right)\\
   {\cal V} (\phi ) &= \frac{m^2}{2\beta^2} \left(e^{\beta\phi} + e^{-\beta\phi }\right)
\end{align}
where $b_0, b_1$ are arbitrary real constants. This result was first discovered in the
case of the sine-Gordon model by Ghoshal and Zamolodchikov \cite{gz}). It is interesting to
notice that each term in the boundary potential ${\cal B}$ is, up to a constant, the square root of the
corresponding term in the bulk potential ${\cal V}$. This particular feature turns out to be universal
within the class of affine Toda field theory models once boundaries are incorporated \cite{cdrs, bcdr},
although the number of free parameters introduced via a boundary condition is generally much
more severely restricted than it is for the sine/sinh-Gordon model. In fact, these constraints on
boundary parameters are somewhat mysterious. For more information on the sinh-Gordon
case, including a discussion
of specific solutions, see \cite{cdr}. For further information on the behaviour of
classical soliton solutions in the sine-Gordon and other models with a boundary consult \cite{b1}.
It is also possible to render the boundary dynamical in a consistent manner. To learn about one
such possibility, consult \cite{bd}.

In the quantum field theory we envisage a situation where a particle travels towards the boundary,
hits it, and bounces back. Then, the incoming particle state will be related to the
outgoing particle state by reversing momentum and by a phase factor (the `reflection' factor).
Thus,
\begin{equation*}
 |k\rangle_{\rm out}=R(k,b_1, b_2)|-k\rangle_{\rm in}.
\end{equation*}
One question to ask concerns the relationship between the $S$-matrix, which describes the
scattering of two sinh-Gordon particles, and the reflection factor $R$. This is known for
sinh (or sine)-Gordon but remains an open question for almost all the other integrable
field theories with boundary. Thus, for the sinh-Gordon model the $S$-matrix is given
by the rapidity-dependent phase factor
\begin{equation}\label{shGS}
S=- \frac{1}{(B)_\Theta(2-B)_\Theta},
\end{equation}
where the notation conceals a certain complexity:
\begin{equation}\label{bracketdef}
(x)_\Theta = \frac{\sinh\left(\frac{\Theta}{2}+\frac{i\pi x}{4}\right)}
                  {\sinh\left(\frac{\Theta}{2}-\frac{i\pi x}{4}\right)},
\end{equation}
and
\begin{equation*}
B(\beta)= \frac{2\beta^2}{8\pi + \beta^2},
\end{equation*}
with $\Theta=\theta_1-\theta_2$, the difference in rapidities of the two particles
($\theta_1>\theta_2$ and $k_i=m\sinh\theta_i, \ i=1,2$). On the other hand, using the same
notation, the reflection factor for just one particle depends on the two boundary
parameters as well as the bulk coupling and the rapidity $\theta$ of the particle:
\begin{equation}\label{shGR}
R=\frac{(1)_\theta(1+B/2)_\theta (2-B/2)_\theta}{(1+E)_\theta(1-E)_\theta(1+F)_\theta(1-F)_\theta}
\end{equation}
where
\begin{equation*}
E=(a_0+a_1)(1-B/2), \ F=(a_0-a_1)(1-B/2), \ b_j=\cos a_j\pi , \ j=0,1.
\end{equation*}
For further details concerning this expression, and further references, see \cite{g,c1,cd,ct}

Returning for a moment to the free field situation there is another phenomenon
which is worth examining. Let $k= -i\lambda, \  (\lambda <0)$. Then, provided $-m<\lambda<0$,
there is a time-periodic solution to the equations \eqref{KGhalfline} which declines exponentially
away from the boundary. In more detail, this is given by
\begin{equation*}
\phi= A \cos\omega t \, e^{-\lambda x}, \ \omega^2=m^2-\lambda^2,
\end{equation*}
and represents a `boundary bound state'. It is clearly a result of the competition between the
bulk energy (always positive) and the boundary energy (negative if $\lambda <0$) and, after
quantisation leads to a tower of states equally spaced in energy, actually very similar to
the spectrum
of a harmonic oscillator. These new kinds of states also exist in the nonlinear models and
have been investigated for the sinh-Gordon model in \cite{cd,ct}. These states illustrate
the much richer content of the nonlinear models once a boundary is introduced.

\section{Two boundaries}

Once it becomes possible to deal with a single boundary, it is natural to wonder about
two of a similar type. Once again it is instructive to look at the free field first,
supposing it is defined on the interval $-L<x<L$ with suitable boundary conditions
at each end:
   \begin{eqnarray*}
 (\partial ^2 + m^2)\phi =0 , &\ \  \text{$(|x| < L)$}\\
       \partial _x\phi = \mp \lambda _{\pm }\phi , &\ \ \ \text{$(x = \pm L)$},
   \end{eqnarray*}
where $\lambda_\pm$ are two real parameters. As before, it is appropriate to
consider a solution with a specific frequency $\omega$,\ $\omega^2=m^2+k^2$ and impose the pair
of conditions at the ends of the interval. A little algebra reveals a nice factorised form
for a relationship which in effect determines the possible frequencies the field might adopt
within the interval. Explicitly, the equation determining the frequencies (via $k$) is
\begin{equation}\label{KGtwoboundary}
e^{4ikL} = R_+(k)R_-(k)
\end{equation}
where $R_\pm(k)$ are the reflection factors for the reflections at the two ends. Effectively,
each boundary works independently of the other.

One important question is whether a similar story will be the case
for the nonlinear models such as the sinh-Gordon model. In that
case, even classically, the solutions within an interval are not
known explicitly, and it has not yet been shown that a similar
factorisation to that of \eqref{KGtwoboundary} will occur in all
cases. In fact, to make progress on this it seems likely to
require a better understanding of the `$N$-zone' solutions
discovered some years ago by Mumford \cite{m} and independently by
Dubrovin and Natanzon \cite{dn}. These are constructed from
$\vartheta$-functions and make essential use of the Fay
identities, which appear to fit very naturally into this context.
Nevertheless, incorporating the boundary conditions has proved
elusive so far. This aspect of the exact solutions is decidedly
geometrical in flavour because the relationship between the
$\vartheta$-functions and Riemann surfaces plays a central role,
and would probably repay closer scrutiny.

\section{Defects}

Another possibility, which has been explored in \cite{bcz1,bcz2}, introduces an internal boundary
or `defect' at which a field may have a discontinuity, or at which two fields of differing
character might meet. Actually, `defects' are a common phenomenon in many areas
of physics; one only has to think of a `bore' or `hydraulic jump' at which the level
of a fluid flow suddenly changes, a `shock' front at which fluid flow suddenly changes
from subsonic to supersonic, or a dislocation within a crystal or
other material. In all cases, there is a discontinuity in some physical quantity while
others may remain continuous. However, in the context of these lectures the interest is
in integrability rather than any specific phenomenon.
There is nothing special about one defect, any number of them might be allowed
sprinkled along the real line at
$x_1\ x_2,\ \dots$, but, for a first look, consider just one at $x=0$, and
two scalar fields.
The field in the region $x<0$ is denoted by $\phi $ and the field in the
region $x>0$ by
$\psi $.

A suitable Lagrangian for the pair of scalar fields $\phi ,\psi
$ is;
\begin{align}\label{defectlagrangian}
\nonumber \mathcal{L} &= \theta (-x)\left\{ \frac{1}{2}(\partial \phi )^2-{\cal V}(\phi )
\right\}+\theta (x)\left\{\frac{1}{2}(\partial \psi )^2-{\cal W}(\psi )\right\}\\
             &\qquad \qquad +\delta (x)\left\{\frac{1}{2}(\phi \dot{\psi}-\psi \dot{\phi})-
             {\cal B}(\phi ,\psi )\right\}
\end{align}
in which $B(\phi ,\psi )$ represents the defect potential. Note, this expression
has not been derived from any specific physical system; it is a purely theoretical
possibility which might (or might not) have any real relevance. The field equations
are as follows
\begin{eqnarray*}
 \left\{ \begin{array}{ll}
     \partial ^2\phi =\displaystyle{-\frac{\partial {\cal V}}{\partial \phi }}  & (x < 0) \\
     \partial ^2\psi =\displaystyle{-\frac{\partial {\cal W}}{\partial \psi }}  & (x > 0)
\end{array} \right.
\end{eqnarray*}
with boundary conditions at $x=0$:
\begin{equation}\label{defectboundary}
  \begin{cases}
      \partial _x\phi =\partial _t\psi -\displaystyle{\frac{\partial B}{\partial \phi }}\\
      \partial _x\psi =\partial _t\phi +\displaystyle{\frac{\partial B}{\partial \psi }}
  \end{cases}
\end{equation}
The particular form of \eqref{defectlagrangian} is really required by integrability and the
lack of time reversal invariance is a special feature.

However, since time is limited, and the integrability will not be described, at least not
with any details, it is nevertheless
worth picking up on a particular point which already captures much of the detail which
integrability would imply. For further information consult \cite{bcz1,bcz2,cz}.

Consider the momentum carried by the two fields in their respective domains. This is given by the
expression
\[
P = \int _{-\infty }^0dx\, \partial _t\phi \partial _x\phi +\int _0^{\infty }dx\, \partial _t\psi
\partial _x\psi \]
and is not expected to be conserved because translation invariance is explicitly broken by
placing a defect at $x=0$. However, using the equations of motion in the two domains, and
integrating by parts in the usual way, leads to
\[
\begin{split}
  \dot{P}&= \int _{-\infty }^0dx\, \partial _x\left(\frac{1}{2}(\partial _x\psi )^2-{\cal V}
         + \frac{1}{2}(\partial _t\psi )^2\right)\\
          &\qquad \qquad \qquad \qquad \qquad  +\int _0^{\infty }dx\, \partial _x
              \left(\frac{1}{2}(\partial _x\psi )^2-{\cal W}+\frac{1}{2}(\partial _t\psi )^2\right)\\
\newline
        &=\left[\frac{1}{2}(\partial _x\phi )^2+\frac{1}{2}(\partial _t\phi )^2-{\cal V}\right]_{x=0}
           -\left[\frac{1}{2}(\partial _x\psi )^2 + \frac{1}{2}(\partial _t\psi )^2 -
           {\cal W}\right]_{x=0}
\end{split}
\]
Using the boundary conditions (\eqref{defectboundary}, the latter can be rearranged to
\begin{align}\label{Pdot}
 \dot{P} &=\left[-\dot{\psi }\frac{\partial {\cal B}}{\partial \phi }-\dot{\phi }
 \frac{\partial {\cal B}}{\partial \psi }
   +\frac{1}{2}\left(\frac{\partial B}{\partial \phi}\right) ^2 -\frac{1}{2}
   \left( \frac{\partial B}{\partial \psi }\right)^2
    -\left( {\cal V}(\phi )-{\cal W}(\psi )\right) \right]_{x=0}\nonumber\\
    &= -\frac{\partial U}{\partial t}
\end{align}
\noindent
where $U$ is a functional of the fields evaluated at $x=0$
provided the following two equations hold:
\begin{align*}
 &\frac{\partial^2{\cal B}}{\partial \phi ^2}=\frac{\partial^2 {\cal B}}{\partial \psi^2}, \\
  &\frac{1}{2}\left\{ \left(\frac{\partial{\cal B}}{\partial \phi }\right)^2
              -\left(\frac{\partial {\cal B}}{\partial \psi }\right)^2\right\} ={\cal V}(\phi )-{\cal W}(\psi ).
\end{align*}
The first of these is a `wave' equation for ${\cal B}$ in terms of the two fields evaluated
at $x=0$; the second is a nonlinear condition constraining the relevant solutions. There are several
possibilities but the simplest to examine is a pair of free fields with equal mass parameters.
Then it is not difficult to check that
\begin{align*}
  m_{\phi }&=m_{\psi }=m, \qquad {\cal B}=\frac{m\lambda }{4}(\phi +\psi )^2+\frac{m}{4\lambda }(\phi -\psi )^2\\
 & {\cal V}(\phi )=\frac{1}{2}m^2\phi ^2,\qquad {\cal W}(\psi )=\frac{1}{2}m^2\psi ^2.
\end{align*}
Note that, generally, $$\displaystyle{\lim_{x \rightarrow 0}\phi (x ,t)\ne \lim_{x \rightarrow 0}
\psi (x ,t)}$$ and the fields can be discontinuous at the location of the defect. Nevertheless,
it is perfectly possible to modify slightly the definition of momentum so that the revised momentum
$P+U$ is conserved despite the absence of the usual Noether argument using
translation invariance. In a way this is quite surprising. Note also that there is an additional
real parameter $\lambda$ which is
free and introduced by the defect; in the limit $\lambda\rightarrow 0$ the defect disappears
since the fields to the left and the right of it match exactly in that limit.

With nonlinear fields there is a variety of possibilities \cite{bcz1,bcz2}, but suffice it to say
here that the defect conditions \eqref{defectboundary}, at least in the cases which have been
analysed so far, turn out to be B\"acklund transformations frozen at the location of the defect.
Thus, to take the sine-Gordon model as an example, the defect boundary conditions turn out
to be
\begin{eqnarray}
\partial_x\phi -\partial_t\psi &= -\frac{m\lambda}{\beta}\sin\beta\left(\frac{\phi+\psi}{2}\right)-
\frac{m}{\beta\lambda}\sin\beta\left(\frac{\phi-\psi}{2}\right)\\
\partial_x\psi -\partial_t\phi &= \frac{m\lambda}{\beta}\sin\beta\left(\frac{\phi+\psi}{2}\right)-
\frac{m}{\beta\lambda}\sin\beta\left(\frac{\phi-\psi}{2}\right),
\end{eqnarray}
where $\lambda$ is again a free parameter.
This fact, too, has something of a geometrical flavour about it, given that B\"acklund was
interested in spaces
of constant negative curvature at the time he developed the transformation bearing his name
\cite{b}. Also,
the solitons of the sine-Gordon model are normally transmitted by a defect - but delayed and
possibly converted to anti-solitons - or they may be absorbed. All these possibilities are allowed
because the defect can `store' energy/momentum and topological charge. These facts allow the
intriguing possibility of controlling solitons (see \cite{cz} for some further ideas - an
article which appeared after these talks were given).

From the point of view of the quantum field theory, these ideas are not yet fully developed
although there is some literature treating aspects of the story
(see for example \cite{dms,kl,cf,mrs}).

\section{Generalised Lax pairs}

It is not the purpose here to give a full description of classical integrability. For that, the
interested reader is recommended to consult one of the books on the subject, for example \cite{l, ft}.
Instead, a few words will be said concerning the specific situations mentioned in the previous
sections.

There are several routes to demonstrate the classical integrability in the variety of cases mentioned
above. One is to systematically check conserved quantities, starting with those of lowest spin
(for example, the sine/sinh-Gordon model in the bulk has conserved quantities for each odd spin
and it is enough to check the conservation of a modified form of the `energy-like' combination
of spin three charges in the presence of a boundary \cite{gz}).
However, this is a painstaking procedure
and it is better to develop a method which generates all conserved quantities simultaneously.
The generalised Lax pair provides such a method adaptable to either the boundary
or the defect situations.

The formulation of integrability in the presence of a boundary has its origins in the
pioneering work of Sklyanin \cite{s}, and it was adapted more recently to accommodate
the sine(sinh)-Gordon model with a boundary by MacIntyre \cite{m1}.
However, the procedure which will be described here
is somewhat different in style though allowing a direct computation of the crucial element of
Sklyanin's formulation, which otherwise has to be calculated using compatibility with
the classical version of the Yang-Baxter relations. Time does not permit a detailed
comparison of all these ideas, unfortunately.

The Lax pair idea can be generalised to the full set of affine Toda field theories (for a
classification, see \cite{mop, ot}), which are defined conveniently in terms of
Lie algebra data as follows. There is a set of scalar fields $\phi_a,\ a=1,2,\dots,r$
where $r$ is the rank of a Lie algebra $g$ whose interactions are described by the
Lagrangian density
\begin{equation}\label{Todalagrangian}
{\cal L}=\frac{1}{2}\partial_\mu\phi\cdot\partial^\mu\phi - \frac{m^2}{\beta^2}
\sum_{i=0}^r\, n_i\, e^{\beta\alpha_i\cdot\phi},
\end{equation}
where $\alpha_i,\ i=1,2,\dots,r$ is a set of simple roots for $g$, $m$ is a mass scale and
$\beta$ is the bulk coupling constant. The vector $\alpha_0$, defined by
$$\alpha_0=-\sum_{i=1}^r n_i \alpha_i$$
is the Euclidean part of the additional (affine) root in the Ka\v c
classification of the affine root systems (by convention the long
simple roots in any root system are taken to have length $\sqrt{2}$). The integers
$n_i,\ i=1,2,\dots,r$ are characteristic of a particular root system and $n_0=1$.
Thus, taking the simplest example of a root system corresponding to $a_1$, one
takes $\alpha_1=\sqrt{2}=-\alpha_0$ and, apart from a rescaling of $\beta$,
\eqref{Todalagrangian} is identical to the Lagrangian for the sinh-Gordon model.
In other words, the sinh-Gordon model is simply part of a large collection of
models each possessing the property of integrability (for
 further references see \cite{c}). It is also worth remarking that if the term
in \eqref{Todalagrangian} corresponding to $i=0$ is deleted then what remains
is actually a conformal field theory and of considerable interest in its own right.

The field equations following from \eqref{Todalagrangian} are
\begin{align}\label{Todafieldequations}
 \partial^2 \phi = -\frac{m^2}{\beta}\sum_{i=0}^r\, n_i\alpha_i\, e^{\beta\alpha_i\cdot\phi}
\end{align}
and these may be cast into a Lax pair form making use of (some of) the generators of the
Lie algebra $g$. To see how to do this, consider defining a two dimensional gauge
field $a_t,\ a_x$ as follows:

\begin{align}\label{TodaLaxpair}
\nonumber a_t &= \frac{1}{2} {\bf H} \cdot \partial_x \phi
    + \sum_{i=1}^r m_i
       (\lambda E_{\alpha_i} - \frac{1}{\lambda}E_{-\alpha_i})
       e^{\alpha_i\cdot\phi/2} \\
 a_x &= \frac{1}{2}{\bf H} \cdot \partial_t \phi
    + \sum_{i=1}^r m_i
       (\lambda E_{\alpha_i} + \frac{1}{\lambda}E_{-\alpha_i})
       e^{\alpha_i \cdot\phi/2}.
\end{align}
In \eqref{TodaLaxpair}, ${\bf H}$ is the Cartan subalgebra of $g$ and $E_{\pm\alpha_i}$
are the step operators corresponding to the simple roots, their negatives, or to
$\pm\alpha_0$. The important particular property which allows the Lax pair to work is the fact
that the step operators for simple roots satisfy the following
\begin{equation*}
\left[E_{\alpha_i},\ E_{-\alpha_j}\right]=0,\quad i\ne j =0,2,\dots,r
\end{equation*}
together with the basic commutation relations
\begin{equation}\label{Liealgebra}
\left[{\bf H},\ E_{\pm\alpha_i}\right]=\pm \alpha_i\, E_{\pm\alpha_i},\quad \left[E_{\alpha_i}
,\ E_{-\alpha_i}\right]=\frac{2\alpha_i}{\alpha_i^2}\cdot{\bf H}.
\end{equation}
Armed with these, and a suitable choice of $m_i$ (it is an exercise for you to work it out), the
`curvature' built from $a_t,\ a_x$, that is
\[
 F_{tx}=\partial_t a_x - \partial_x a_t + [a_t, a_x],
\]
vanishes if and only if the field equations \eqref{Todafieldequations} hold
independently of the value of the spectral parameter $\lambda$.

The Lax pair may be used to generate conserved quantities (see \cite{ot})
since the quantity $Q(\lambda)$ defined by parallel transport over the range
$-\infty<x < \infty$,
\[
 Q = {\rm tr} \left(P \exp \int_{-\infty}^\infty dx\, a_x\right),
\]
is conserved automatically as a consequence of the zero curvature condition. Its coefficients
in an expansion in powers of $\lambda$ are individually conserved (and indeed are
in involution). The details of all this are independently interesting but cannot
be pursued here. Instead, a way to develop this idea when there is a boundary
(or a defect) will be sketched next.

\section{Lax pair with boundary conditions}

One needs to proceed slightly differently according to the context.
For field theories with a boundary one way to develop this theme is along the lines of reference
\cite{bcdr}; on the other hand, for
field theories with a defect appropriate references would be \cite{bcz1,bcz2}.
In this lecture, the boundary case will be considered in detail, really to give a flavour of the
kind of mathematics involved and to show how constraints on the boundary potential arise. Other
cases will not be covered in any detail at all.

For theories with a boundary, it was found to be convenient to
construct a
field theory on two overlapping halves of the $x$-axis: $R_{\pm}$
defined as follows.
The half-line $R_-$ consists of the portion $-\infty <x\le b$ and
the half-line $R_+$ is the
portion $a\le x<\infty$, where $a<0<b$. Clearly the two portions
overlap on the region $[a,b]$.
The field in $x\ge b$ is defined in terms of the field in $x\le a$
via a reflection principle:
\begin{equation}\label{fieldreflection}
\phi(x)=\phi (a+b-x), \quad x\ge b.
\end{equation}
In the case of a defect, no such reflection principle would be imposed, of course.

The next step is to modify the components of two-dimensional gauge gauge field entering
the Lax pair by setting in the two overlapping regions
\begin{align*}
R_-:\quad \hat{a}_t^- &= a_t -\frac{1}{2} \theta(x-a)(\partial_x \phi +
\frac{\partial {\cal B}}{\partial \phi}) \cdot {\bf H},\\
 \hat{a}_x ^-&= \theta(a-x)a_x,\\
R_+:\quad \hat{a}_t^+ &= a_t -\frac{1}{2} \theta(b-x)(\partial_x \phi -
\frac{\partial {\cal B}}{\partial \phi}) \cdot {\bf H},\\
 \hat{a}_x^+ &= \theta(x-b)a_x,
\end{align*}
where $\theta(x)$ is the usual step function. It is left as an exercise to
verify that these do indeed constitute a Lax pair whose zero curvature condition
supplies not only the field equations but also the boundary conditions in the two
separate regions. However, on the overlap it is clear $\hat{a}_x^\pm$ vanish
identically, by design, and therefore zero curvature requires each of
the components $\hat{a}_t^\pm$ to be constant (ie independent of $x$), but not necessarily equal.
To put it another way, these two quantities need not be equal provided they are related to each other
 by a gauge transformation.
In other words, there should be a group element ${\cal K}$, possibly depending upon $t$, with the
property
\begin{equation}\label{Kgauge}
\partial_t{\cal K}={\cal K}\hat{a}^+_t- \hat{a}^-_t{\cal K}, \quad a\le x\le b.
\end{equation}
Then, provided this is the case, the quantity
\begin{align}\label{QKQ}
 Q &= {\rm tr} \left(P \exp \left\{\int_{-\infty}^a dx a_x^- \right\}\, {\cal K}\, P
 \exp\left\{ \int_b^{\infty}dx\, a^+_x\right\}\right),
\end{align}
will be conserved. Because of the reflection principle, the part involving $a_x^+$
can be reinterpreted as a parallel transport back along the left half-line. In fact,
the expression \eqref{QKQ} is the starting point for Sklyanin's analysis \cite{s}.
However, following \cite{bcdr} instead, and making a mild assumption,
 it is possible to calculate ${\cal K}$ and in the process determine the boundary
 potential ${\cal B}$.

 Suppose ${\cal K}$ does not depend on the  fields, and suppose further that
\[
 \partial_0 {\cal K} = 0.
\]
At first sight these appear to be strong assumptions. However, an alternative,
independent approach, investigating the low spin conserved charges, leads to
precisely the same conclusions. For that reason these assumptions are actually
quite mild and the boundary potential derived from them seems to be rather
general.

Making these two assumptions  and using the explicit expressions
for the two gauge components $a_t^\pm$, equation \eqref{Kgauge} becomes the following
\begin{equation}\label{Kalgebra}
 \frac{1}{2}\left[{\cal K}, \ \frac{\partial {\cal B}}{\partial \phi}\cdot{\bf H}\right]_+
 = -\left[{\cal K},\ \sum_{i=0}^r m_i \left(\lambda E_{\alpha_i} -
 \frac{1}{\lambda}E_{-\alpha_i}\right) e^{\alpha_i \cdot \phi/2}\right]_-.
\end{equation}
It is worth noting, first of all, that there is an anti-commutator on the left hand side
and a commutator on the right, and secondly, that although ${\cal K}$ depends upon the
spectral parameter $\lambda$, the boundary potential ${\cal B}$ and of course the
fields $\phi$  do not. These facts are powerful properties of equation \eqref{Kalgebra}.

First of all, if ${\cal K} =1 $ the commutator on the right hand side of \eqref{Kalgebra} vanishes
identically, while the anti-commutator on the left hand side vanishes only provided
\[
\frac{\partial {\cal B}}{\partial \phi_a} =0.
\]
Thus ${\cal K}=1$ is equivalent to the Neumann condition
$$ \partial_x\phi_a=0.$$

On the other hand, suppose ${\cal K}$ is well-defined at $\lambda =0$. Then
${\cal K}(0)$ will have to commute with all
$E_{-\alpha_i}$, otherwise the second term on the right hand side of \eqref{Kalgebra}
would make no sense. Hence, ${\cal K}(0)$ is a central element of the group and,
since \eqref{Kalgebra} is linear in ${\cal K}$, one might as well take ${\cal K}(0) = 1$.
In that case, the group element ${\cal K}$ should have an expansion of the form
\[
{\cal K} = e^{\sum_{n=1}^\infty \lambda^n k_n}.
\]
Using this, equation \eqref{Kalgebra} can be solved iteratively.

For the simplest case $g=a_1$, it is convenient to take $\alpha_1=\alpha=-\alpha_0$
and proceed directly to find
\[
{\cal K}(\lambda)=I + \frac{\lambda}{1-\lambda^4}\left( \begin{array}{cc} 0 & b_1-\lambda^2b_0\\
                 b_0-\lambda^2b_1  & 0
       \end{array}\right)
\]
with the corresponding boundary potential given by
\[
 B = b_1 e^{\alpha \phi/2} + b_0 e^{-\alpha\phi/2}.
\]
This is, up to rescalings by the bulk coupling, in  agreement with
\eqref{sinhGpotentials}. Hopefully, all the signs and so on are correct in these
expressions but, in any case, it is an exercise to check it! To do so, it is
helpful to use the basis
\begin{equation*}
H=\frac{\alpha}{2}\left(%
\begin{array}{cc}
  1& \phantom{-}0 \\
  0 & -1 \\
\end{array}%
\right)
\quad
E_\alpha =\left(%
\begin{array}{cc}
  0 & 1 \\
  0 & 0 \\
\end{array}%
\right)
\quad
E_{-\alpha} =\left(%
\begin{array}{cc}
  0 & 0 \\
  1 & 0 \\
\end{array}%
\right)
\quad \alpha^2=2.
\end{equation*}

Returning to the general case consider the terms of $O(1)$. Balancing these on
both sides of \eqref{Kalgebra} gives
\[
 \frac{\partial B}{\partial \phi} \cdot {\bf H}
 = \left[ k_1, \sum_{i=0}^r\, m_i\, E_{-\alpha_i} \, e^{{\alpha_i \cdot \phi/2}}\right]_-,
\]
where $m_i^2 = n_i \alpha_i^2/8$.
Using the Lie algebra commutation relations \eqref{Liealgebra} it is clear the only solution
to this must be to take
\begin{equation}\label{kone}
 k_1 = \sum_{i=0}^r c_i E_{\alpha_i},
\end{equation}
where $c_i,\ i=0,1,\dots ,r,$ are a set of constants, and
\[
 \frac{\partial {\cal B}}{\partial \phi}
  = \sum_{i=0}^r m_i \, \alpha_i\, c_i \, \frac{2}{\alpha_i^2}\, e^{\alpha_i \cdot \phi/2}.
\]
Thus, on integrating, the characteristic expression for the boundary potential follows.
\[
{\cal  B} = \sum_{i=0}^r b_i e^{{\alpha}_i \cdot{\phi}/2}
\]
with $b_i= \sqrt{2n_i/\alpha_i^2}\,\, c_i$. At this stage it appears as though there are
$r+1$ free parameters $b_i$. However, there are surprises yet to come.

At the next order, $O(\lambda)$, it is not hard to check that $k_2\equiv 0$. (You should
discover that $k_2$ has to commute with $E_{-\alpha_i},\ i=0,\dots ,r$ and therefore
contributes a factor in ${\cal K}$ which commutes with everything else. Since, \eqref{Kalgebra}
is homogeneous in ${\cal K}$ any such factor can be scaled out, effectively setting
$k_2$ to zero.) However, that
is not yet the end of the story because at $O(\lambda^2)$ there is a serious-looking equation
for $k_3$ in terms of $k_1$. In detail, it is
\begin{align}\label{kthree}
\left[ k_3, \  m_i E_{-\alpha_i} \right]_-
 =\left[ k_1, \  m_i E_{\alpha_i}  +
 \frac{b_i}{24} \left[ k_1, \alpha_i
 \cdot {\bf H} \right]_-\right]_-, \quad i=0,1,\dots,r.
\end{align}
A case by case analysis of \eqref{kthree} is provided in \cite{bcdr} and there are
significant differences between the different choices of Lie algebra data. One
important class of root systems is `simply-laced', meaning that each root of
the Lie algebra has the same length (conventionally taken to be $\sqrt{2}$) and
for this class, the consequences of \eqref{kthree} are relatively simple and striking.

Given that $k_1$ according to eq\eqref{kone} is composed of step operators corresponding to
level one roots (the simple roots and, modulo the Coxeter number, $\alpha_0$),
and given that the levels must balance across the equation \eqref{kthree},
it should be clear that $k_3$ ought to have the form
\[
 k_3 = \sum_{\text{level 3 roots $\beta$}} d_{\beta}\  E_\beta,
\]
where the coefficients $d_\beta$ are to be determined. Using the expressions for
${\cal B}$ and $k_1$ previously obtained leads not only to expressions for
$d_\beta$ but also further constraints on the coefficients $b_i,\ i=0,\dots, r$.
In fact, for the simply-laced cases, one discovers $b_i^2 =4n_i$ for every $i=0,
\dots, r$ - with
the only exception being the simplest case $a_1$ (which has no level three roots anyhow).
This curious fact was first pointed out for the $a_n$ series of cases in \cite{cdrs}
by examining the low spin conserved quantities.
To see why this is so, evaluate \eqref{kthree} carefully using the commutation
relations of the Lie algebra to obtain
\begin{align}\label{k3again}
\nonumber \sum_{\beta}d_\beta m_i \epsilon(\beta, -\alpha_i)E_{\beta -\alpha_i}
&=\sum_{j} c_j m_i\epsilon(\alpha_j,\alpha_i)E_{\alpha_j+\alpha_i}\\
&\quad -\sum_{k\ne l}\frac{b_i bc_k c_l}{24}\alpha_i\cdot\alpha_l \epsilon(\alpha_k, \alpha_l)
E_{\alpha_k +\alpha_l}.
\end{align}
Now, for a simply-laced root system $E_{\alpha_j+\alpha_i}$ cannot appear on the
left hand side,
since that would require $\alpha_j+2\alpha_i$ to be a root, which it cannot be.
Therefore,
the coefficient of $E_{\alpha_j+\alpha_i}$ must also vanish on the right hand side, which
requires
$$ m_i+\frac{b_i c_i}{24}\, (\alpha_i\cdot\alpha_j - \alpha_i\cdot\alpha_i)=0.$$
However, the vector $\alpha_j+\alpha_i$ is only a root when $\alpha_i$ and $\alpha_j$
are adjacent
on the Dynkin diagram for $g$, and then $\alpha_i\cdot\alpha_j=-1$. Hence, $b_ic_i=8m_i$
or, $b_i^2=4n_i$,
for each $i=0,1,\dots, r$.

In other words for each of the cases $a_r,\ (r\ge 2),\ d_r, \ (r\ge 4),\ e_r, (r=6,7,8)$
the possible boundary potentials for which integrability is maintained consist of a discrete
set ($2^{r+1}$ choices) with no free parameters at all. This is very surprising. One
might have thought a priori that the more complex the root system the more freedom there
might be. In fact, it is only in some of the non-simply-laced cases where some free parameters
remain. However, even there, the set of possibilities is severely restricted.

\section{Brief discussion}

In just two lectures it is difficult to do justice to a topic which by now has a sizeable
literature and which is interesting within mathematical physics both from a classical and
a quantum field theoretical point of view. Moreover, although there has been  interest in Toda
theory within the geometry
community, largely because of the relationship with harmonic maps, it is not yet clear
to what extent the phenomena discussed briefly here will find a geometrical context.
Unfortunately, there is also the problem of language! Nevertheless, a possible starting
point might be the book by Guest \cite{guest}. There is  a
vast literature on the subject of harmonic maps and perhaps, somewhere, there is a natural
home for the affine Toda field theories with one or two boundaries.

\section{Acknowledgements}

I am grateful to the organisers of the meeting for asking me to
air these ideas. Some parts of the lectures were given previously
at the EUCLID/E\"otv\"os School held at the Bolyai College,
Budapest, under the auspices of the European Training Network
EUCLID (contract number HPRN - CT - 2002-00325) within which both
the mathematical physics groups of the Universities of Durham and
York are partners. I am indebted to many colleagues, especially
Peter Bowcock, Patrick Dorey, Gustav Delius, Ryu Sasaki, Evgeny
Sklyanin, Anne Taormina and Cristina Zambon, for discussions about
these matters. I am also grateful to the British Council and the
Japanese Society for the Promotion of Science, and to the Yukawa
Institute for Theoretical Physics, University of Kyoto, for
hospitality while these notes were being prepared.

\end{document}